\documentclass[pre,amsmath]{revtex4}  
\usepackage{graphicx}

\begin{document}

\title{Dynamic Exploration of Networks: from general principles to the traceroute process}

\author{Luca Dall'Asta} 
\affiliation{Abdus Salam International Center for Theoretical Physics, Strada Costiera 11, 34014, Trieste (Italy)} 


\begin{abstract}
Dynamical processes taking place on real networks define on them evolving subnetworks whose topology is not necessarily the same of the underlying one. We investigate the problem of determining the emerging degree distribution, focusing on a class of tree-like processes, such as those used to explore the Internet's topology. A general theory based on mean-field arguments is proposed, both for single-source and multiple-source cases, and applied to the specific example of the traceroute exploration of networks. Our results provide a qualitative improvement in the understanding of dynamical sampling and of the interplay between dynamics and topology in large networks like the Internet. 
\end{abstract}

\maketitle

\section{Introduction}\label{sec1} 
The theory of {\em complex networks} is a primary tool in the interdisciplinary research in complex systems.  It provides a simple (coarse-grained) description of natural as well as social and technological complex systems, by means of mapping their elements and interactions on the vertices and the edges of a graph \cite{bornholdt}. Different systems correspond to microscopically different networks, but their general statistical properties may display similarities. 
Using the tools of statistical physics in analyzing experimental data, researchers have obtained important insights on the structure of these complex networks, discovering universal properties that are common to very different systems. In particular, most of real networks seem to be sparse, with a very small diameter (small-world property), and to present large fluctuations in the connectivity (degree) of the nodes. \\
The theoretical models pretending to explain the origins of these universal properties can be roughly divided in two classes: static models, in which the network structure is the result of some precise combinatorial assignment, and growing models, in which new nodes attach to a pre-existing network by means of a given number of links \cite{dor-book,bar-review}.  
Much less attention was instead devoted to  the other very important issue concerning the relation between networks structure and the processes used to collect the experimental data. Some questions arise naturally: are the experimental data really reliable? can the ubiquity of some properties be the result of systematic biases introduced by the sampling process itself?\\
These questions have been recently addressed in a series of works \cite{delos,stumpf,clauset1,lee,dallasta} that study the sampling of networks from both  static and dynamical points of view. Among them, the work by Petermann and De Los Rios \cite{delos} is the first in which the idea that different generation algorithms may lead to different topological properties is exploited to show analytically that the observed power-law degree distributions can be an artifact of biases affecting the sampling techniques. \\
This issue is particularly relevant for the physical Internet, that is the prototype of a complex network: a huge number of computers are connected together to form a continuously evolving, self-organizing network, on which a large variety of dynamical processes take place \cite{ves-book}. As the robustness and performances of large technological networks strongly depend on their topology, an accurate, unbiased knowledge of the Internet is necessary to support the work of practitioners in improving its  structure and functioning.\\ 
Motivated by the importance of correctly knowing network's topology, the present work is devoted to clarify the mechanisms governing the main techniques used in the dynamical sampling of complex networks.
In order to accomplish this task, we first present the general framework that applies to a large class of tree-like dynamical processes, then we focus on the {\em traceroute sampling} of networks with homogeneous and heterogeneous degree profiles.  We show that the role played by the dynamics is very different in homogeneous and heterogeneous networks. In the latter case, that should correspond to the case of the Internet, the sampling process introduces some level of inhomogeneity that paradoxically increases the sampling accuracy of the tail of the degree distribution. \\ 
Our results are in agreement with previous analytical and numerical achievements, but allow a more unitary view of the problem. 
Dynamical sampling techniques are examples of the interplay between networks topology and the dynamics occurring on it.  Therefore the dynamical approach presented in this paper could be considered as a theoretical improvement with respect to previous analytical studies based on nodes exposure methods \cite{clauset2,cohen}, that are essentially static algorithms.  \\
Moreover, our method can be easily adapted to study other sampling algorithms expressly designed to acquire ego-centric views of a network, such as network's tomography \cite{tomography,kalisky} and crawling algorithms in community networks (the WWW, P2P networks, etc.) \cite{crawling}. 
Similar dynamical patterns are observed also in network processes that are not directly related to sampling, such as epidemics \cite{epidemic}, rumors spreading \cite{rumors}, diffusion of innovations \cite{innovation} and threshold models \cite{threshold}.\\ 
In summary, we provide a unified description of a class of dynamical phenomena in terms of tree-like evolving structures on networks, and explain how the dynamical rules influence the emerging topological properties, by focusing on the degree distribution. 

The paper is organized as follows. In Section~\ref{sec2}, we present the general theoretical formalism that can be used to study the degree distribution generated by tree-like processes on networks. The important application to the traceroute problem is discussed in Section~\ref{sec3}. Some conclusions are presented in Section~\ref{sec4} together with examples of other possible applications.

\section{General formalism for tree-like processes}\label{sec2}

\begin{figure}[t]
\centerline{
\includegraphics*[width=0.4\textwidth]{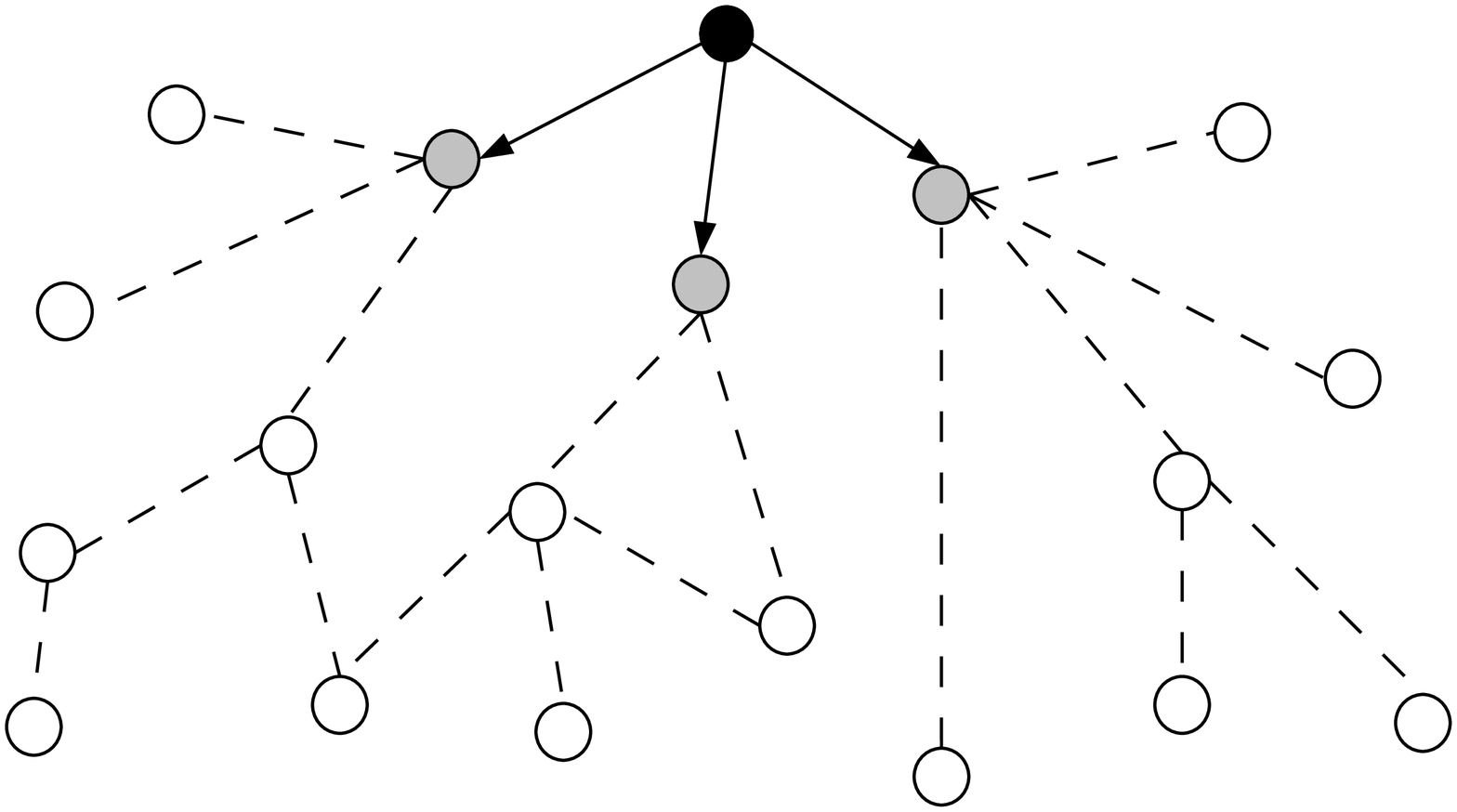}
\includegraphics*[width=0.4\textwidth]{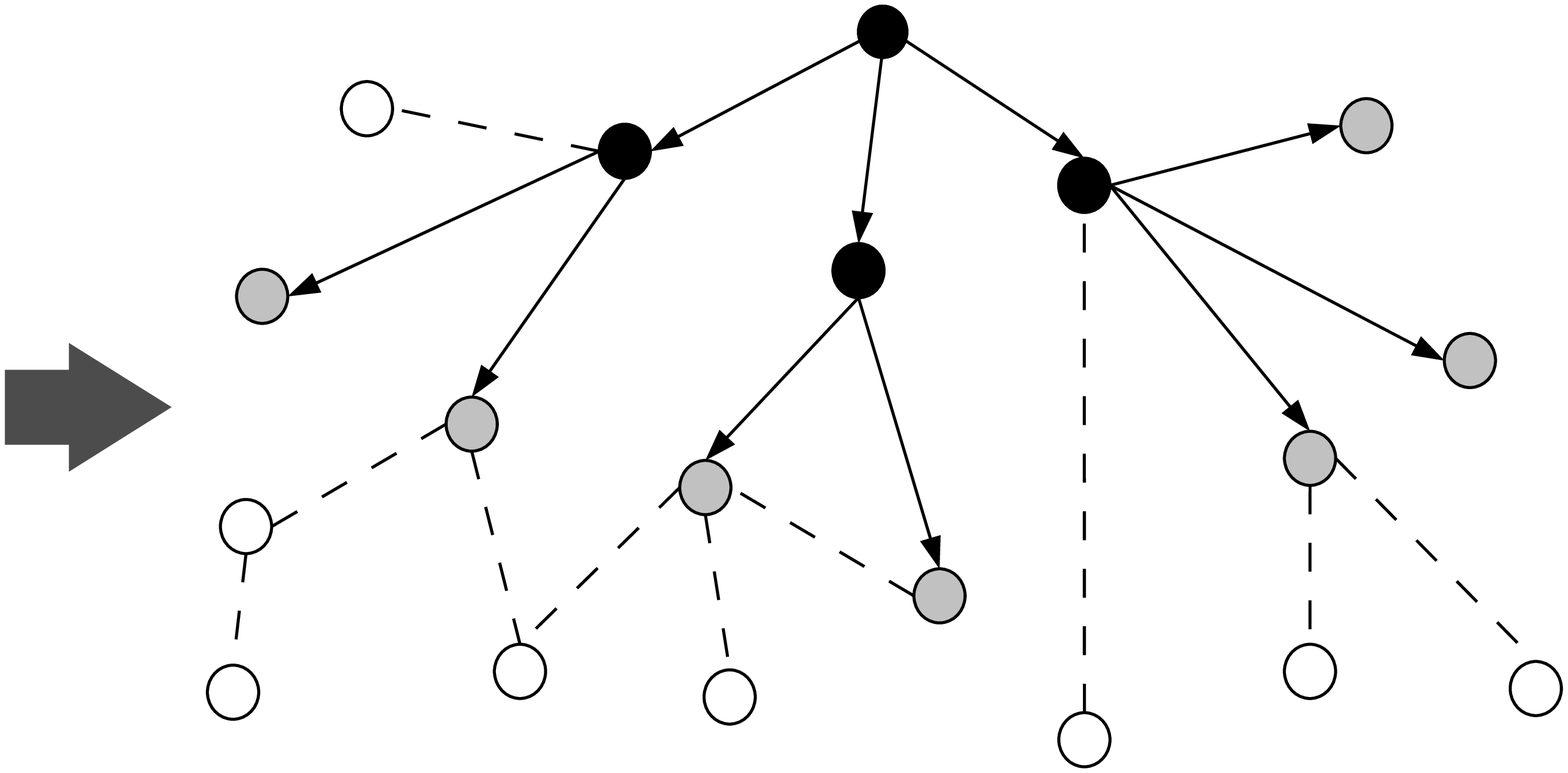}}
\caption{Sketch representing the evolution of the considered class of dynamical processes.  Starting from a single source, the nodes of the network are progressively visited. Once the nodes at the interface (grey nodes) have spread the process towards their still unknown neighbors (white nodes), they are moved to the bulk (black nodes). 
} 
\label{fig0}
\end{figure}

Among the various types of algorithms and dynamical models evolving on networks, we take into account those corresponding to the following general dynamical picture, that is also valid for the special case of traceroute-like explorations.
Let us assume the process starts from a single node and propagates iteratively throughout the network. At each temporal step, some nodes at the interface of the growing cluster are selected and some of their still unreached neighbors are visited. The latter ones become part of the interface, while the former interfacial nodes are moved to the bulk of the cluster (a sketch of the dynamics is reported in Fig.~\ref{fig0}).
During the dynamics we can always identify three distinct classes of nodes: bulk, interfacial, and unreached nodes.
When the above process takes place on a random network, in the limit of large network's size $N$, the overall dynamics is well represented by the temporal evolution of some mean-field quantities, the densities of bulk nodes $b(t)$, interfacial nodes $i(t)$, and unreached nodes $u(t)$. Obviously, $u(t)+i(t)+b(t)=1$ always during the dynamics.\\
On a generic random network, however, the degree is not fixed, the nodes being divided in degree classes. The global densities are replaced by degree-dependent partial densities $b_{k}(t)$,  $i_{k}(t)$, and $u_{k}(t)$. The partial density for unreached nodes of degree $k$ is defined as the fraction of nodes of degree $k$ that are still unreached at time $t$. The normalization relation is $\sum_{k} P(k) u_{k}(t) = u(t)$, where $P(k)$ is the degree distribution of the underlying network. The other quantities are defined similarly. \\ 
More in general, one may be interested in situations in which the network is still maximally random (with degree distribution $P(k)$) but with 
some degree correlations, expressed by the conditional probability $P(k|h)$ that a node of degree $k$ is linked to a node of degree $h$. One can also consider the nodes divided in {\em types} $\alpha$, i.e. discrete or continuous states defined on the nodes. Types are used to create multi-partite networks and to encode some non-topological feature. Another way to account for non-topological properties is that of putting  weights on the links, that depend only on the degrees and the types of the extremities. Link weights may account for dynamical properties of the process, such as the transmissibility of a disease \cite{newman}. The internal structure of the population is taken into account defining degree-dependent and type-dependent partial densities of $u^{\alpha}_{k}(t)$, $i^{\alpha}_{k}(t)$, and $b^{\alpha}_{k}(t)$. The global densities are recovered by averaging over all distributions, i.e. $u(t) = \sum_{\alpha} \mathcal{P}(\alpha) \sum_{k} P(k) u^{\alpha}_{k}(t)$.
In order to simplify the formalism, here on we limit our analysis to single-type undirected random markovian networks \cite{boguna}, that are maximally random graphs completely defined by the degree distribution $P(k)$ and the degree correlations $P(k|h)$.\\
According to this mean-field approximated description of the processes, the temporal evolution of the partial densities satisfies a system of differential equations of the type, 
\begin{equation}\label{gensys}
\left\{ \begin{array}{cl} \frac{d }{d t} u_{k}(t) & = f_{u}(\{u_{h}(t)\},\{i_{h}(t)\},t,\cdots)\\
\quad & \quad\\
\frac{d }{d t} i_{k}(t) & = f_{i}(\{u_{h}(t)\},\{i_{h}(t)\},t,\cdots)\\
\quad & \quad\\
\frac{d }{d t} b_{k}(t) & = f_{b}(\{i_{h}(t)\},t,\cdots)~,\\
 \end{array} \right.
\end{equation}
where the arguments of the functions $f_{x}(\cdot)$ depend on the general form of the dynamics described above. For instance, $f_{b}$ is not expected to depend on unreached nodes, $f_{u}$ on the bulk nodes, etc. The system is generally coupled and non-linear, and admits an explicit solution only in very special cases.\\
In statistical physics and theoretical biology, continuous mean-field dynamical equations are commonly used to study models of  population dynamics. The dynamical picture emerging from these mean-field models allows to understand the qualitative behavior of complex phenomena occurring in real systems.
At the same time, it is worth noting that the rigorous derivation of differential equations for random processes on random graphs has been introduced in the mathematic literature only recently, by Wormald \cite{wormald}, and then applied to several problems, including algorithms for the generation of random graphs with a given degree sequence \cite{molloy} and random k-SAT problems \cite{achlioptas}. Wormald's differential equations method provides a powerful tool to prove rigorous bounds for interesting quantities (e.g. distribution's moments) in discrete-time combinatorial processes.
Here we limit our analysis to a qualitative topological characterization of the emerging degree distribution obtained within a purely mean-field statistical physics approach. However, a rigorous formalization of the present approach is desirable as well.

\subsection{Single-source processes}
The degree distribution $\tilde{P}(k)$ of a subnetwork is related to the degree distribution $P(k)$ of the underlying one by
\begin{equation}
\tilde{P}(k) = \sum_{\ell = k}^{\infty} P(\ell) Q(k|\ell)
\end{equation}
where $P(\ell)$ is the degree distribution, that defines the probability of picking up a node of degree $\ell$ in the original network, and $Q(k|\ell)$ is the conditional probability of observing a node of degree $k$ in the subnetwork if its real degree in the complete network is $\ell$. 
In a dynamical framework, the sampling probability depends on the temporal evolution of the overall process. At the beginning, the neighborhood of interfacial nodes is mainly composed of unreached ones, whereas in the final stage of the dynamics most of the nodes have already been visited. The probabilities $P(\ell)$ and $Q(k|\ell)$ are thus replaced by time-dependent quantities $P_{t}(\ell)$ and $Q_{t}(k|\ell)$, that are defined by the evolution rule of the dynamical processes itself.\\
Using the properties of the dynamics, $P_{t}(\ell)$ can be expressed as the probability of picking up a node of degree $\ell$ among the interfacial nodes of the growing cluster at a time $t$, i.e. $P_{t}(\ell)  =  P(\ell) i_{\ell}(t) / i(t)$. Then, the way in which the neighbors of this node are selected and visited depends strictly on the details of the dynamical model. When the growing cluster has a tree-like structure,  a node of degree $\ell$ has only one incoming edge, and the remaining $\ell-1$ links are used to propagate towards unreached neighbors. Let us call $\rho_{h}$ the probability to visit a neighbor of degree $h$ of an interfacial node of degree $\ell$ (it is a function of $u_{h}(t)$), the conditional  probability $Q_{t}(k|\ell)$ becomes 
\begin{equation}\label{eqnQt}
Q_{t}(k|\ell) = \left( \begin{array}{c} \ell -1 \\ k -1 \end{array} \right) {\left[ \sum_{h} P(h|\ell) \rho_{h}[u_{h}(t)] \right]}^{k-1} {\left[1 -  \sum_{h} P(h|\ell) \rho_{h}[u_{h}(t)] \right]}^{\ell -k}~.
\end{equation}
Putting together these two terms and recalling that the global topology is given by averaging over the whole temporal spectrum,  we obtain the following expression for the degree distribution $\tilde{P}_{1}(k)$ of the tree-like structure emerging from the dynamics,
\begin{equation}\label{P1} 
\begin{split}
\tilde{P}_{1}(k) & =  \sum_{\ell=k}^{\infty} \tilde{P}_{1}(k, \ell) =  \sum_{\ell=k}^{\infty} \frac{1}{T} \int_{0}^{T} P_{t}(\ell)  Q_{t}(k|\ell)~ dt~, \\
\quad &= \sum_{\ell=k}^{\infty} \frac{1}{T} \int_{0}^{T} \frac{P(\ell) i_{\ell}(t)}{i(t)} \left( \begin{array}{c} \ell -1 \\ k -1 \end{array} \right) {\left[ \sum_{h} P(h|\ell) \rho_{h}[u_{h}(t)] \right]}^{k-1} {\left[1 - \sum_{h} P(h|\ell) \rho_{h}[u_{h}(t)] \right]}^{\ell -k} ~ dt~, \end{split} 
\end{equation}
where $T$ is the maximum sampling time.   
In the above equation we have also introduced the joint degree distribution $\tilde{P}_{1}(k, \ell)$ (of observing a node of degree $k$ with  real degree $\ell$), that will be useful in the following. \\
The generalizations to multi-type and weighted networks are straightforward once one has correctly considered the evolution equations for the partial densities of bulk, interfacial and unreached nodes. Again, the use of the Bernoulli sampling technique (binomial probability) in selecting neighboring nodes is justified by the spreading like character of the dynamics considered here. For different classes of dynamical processes, e.g. threshold processes \cite{threshold}, the selection mechanism should be modified. 
\begin{figure}[t]
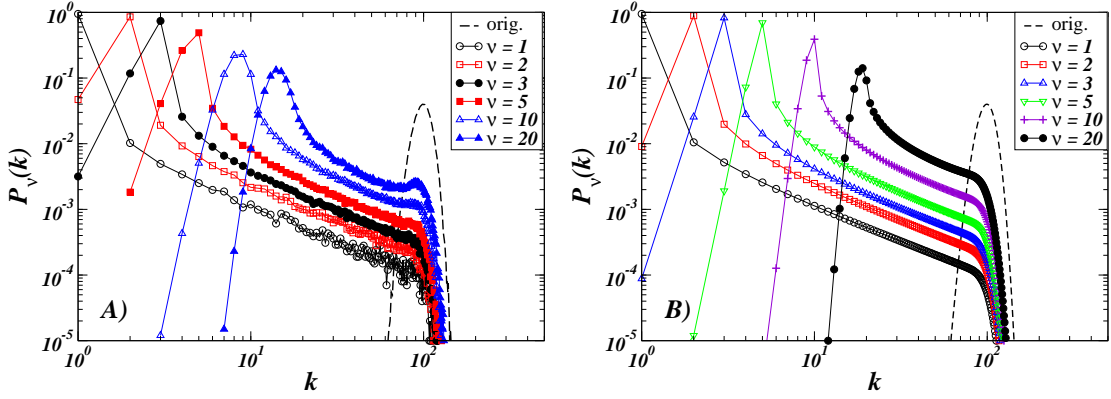

\centerline{
\includegraphics*[width=0.4\textwidth]{fig1A}~~
\includegraphics*[width=0.4\textwidth]{fig1B}
}
\caption{A) Degree distribution $\tilde{P}_{\nu}(k)$ of the network produced by merging together $\nu$ spanning trees generated by one-to-all traceroutes algorithms on a Poissonian random graph of size $N=10^5$ and average degree $z=100$. Increasing the number $\nu$ of sources, i.e. of trees, the degree distribution becomes closer to the original one. A small peak around $z$ is developed because of metric correlations. B) Degree distribution $\tilde{P}_{\nu}(k)$ obtained numerically from the recursion relation in Eq.~\ref{Pnu} for a Poissonian random graph of average degree $z=100$.
} 
\label{fig1}
\end{figure}

\subsection{Multi-source processes}
It frequently happens that several processes are running at the same time, so that the overall measure is obtained merging several single-source structures.
This is indeed the case of Internet's mapping projects, in which single (spanning) trees from different source nodes are merged together in order to get more accurate mappings of the underlying topology.  
A completely different example is provided by the overall infection profile in a population that is object of multiple non-interacting disease spreadings (e.g. e-viruses and worms in P2P communities), each one generating a sort of causal tree.  \\
In order to study multi-source processes, and the emerging degree distribution, we put forward an approximated method based on a simple mean-field argument for the overlap probability of uncorrelated trees. A direct generalization of the single source method presented above is somewhat tricky, since in any node one edge per process is used as an incoming edge and is not available for spreading. Increasing the number of sources one should take care of all possible combinations of these incoming links, that becomes rapidly very complicated. 
Nonetheless, this approach leads to some valuable approximation.
Let us consider a process with two sources, and a node of degree $\ell+1$ for which we fix the incoming edge, so that both trees reach the node from such an edge. Moreover, we assume that the two sampling processes are completely independent and uncorrelated. With these hypotheses, the observed degree distribution, obtained merging two trees is  
\begin{equation}\label{P2_a} 
\begin{split}
\tilde{P}_{2}(k+1) = & \sum_{\ell=k}^{\infty} P(\ell +1) \frac{1}{T^2} \int_{0}^{T} d t_{1} d t_{2}  \frac{i_{\ell +1}(t_{1})}{i(t_{1})} \frac{i_{\ell +1}(t_{2})}{i(t_{2})} \sum_{m,n =0}^{\ell} \sum_{r=0}^{\ell} \left(\begin{array}{c} \ell \\ m \end{array} \right) \left(\begin{array}{c} \ell - m \\ n- r \end{array} \right) \left(\begin{array}{c} m \\ r \end{array} \right) \delta(k-m-n+r) \\
\quad & \quad \times {\left[ \bar{u}(t_{1})\bar{u}(t_{2})\right]}^{r} {\left[\bar{u}(t_{1})\left( 1 -\bar{u}(t_{2}) \right) \right]}^{m-r} {\left[\bar{u}(t_{2})\left( 1 -\bar{u}(t_{1}) \right) \right]}^{n-r}  {\left[\left( 1- \bar{u}(t_{1})\right) \left( 1 -\bar{u}(t_{2}) \right) \right]}^{\ell-m-n+r} ~, 
\end{split}
\end{equation}
where $\bar{u}(t) = \sum_{h} P(h|\ell) \rho_{h}[u_{h}(t)] $ and $\delta(x)$ is the Kronecker's symbol.
Reordering the terms in Eq.~\ref{P2_a}, and using the result for single-source processes, we get 
\begin{equation}\label{P2_b} 
\tilde{P}_{2}(k+1) = \sum_{\ell=k}^{\infty} P(\ell +1) \sum_{m,n,r =0}^{\ell} \mathcal{B}(\ell,m,n,r)  \frac{\tilde{P}_{1}(m+1,\ell +1)}{P(\ell +1)} \frac{\tilde{P}_{1}(n+1,\ell +1)}{P(\ell +1)} \delta(k-m-n+r) ~, 
\end{equation}
in which $\mathcal{B}(\ell,m,n,r)$ is the hypergeometric distribution 
$$\mathcal{B}(\ell,m,n,r)= \left(\begin{array}{c} \ell-m \\ n-r \end{array} \right) \left(\begin{array}{c} m \\ r \end{array} \right) / \left(\begin{array}{c} \ell \\ n \end{array} \right)~.$$
The case in which both trees reach a node through the same edge is obviously very special, as well as Eq.~\ref{P2_a} that does not hold in general. However, one can exploit the picture emerging from Eq.~\ref{P2_b} and generalize it to be valid whatever the choice of the incoming edges. Unlabeling the incoming edges and considering them like the other edges discovered during single-source processes, we get the following approximation for the observed degree distribution in a process with two sources, 
\begin{equation}\label{P2}
\tilde{P}_{2}(k,\ell) \approx P(\ell) \sum_{m,n =1}^{\ell} \sum_{r =0}^{\ell}\mathcal{B}(\ell,m,n,r)  \frac{\tilde{P}_{1}(n,\ell)}{P(\ell)} \frac{\tilde{P}_{1}(m,\ell)}{P(\ell)}  \delta(k-n-m+r)~,
\end{equation}
the sums over $m$ and $n$ start from $1$ since we assume that all nodes are discovered, thus the minimum observed degree is $1$. 
Note that even if two consecutive processes are dynamically uncorrelated, the topological and functional properties of the underlying system always introduce some correlations. For instance, in real networks there are very central nodes that bear a large fraction of the traffic, the so-called {\em backbone nodes}. Depending on the process it may be easier or more difficult to traverse these nodes. Again, in the Internet there are administrative policies governing local routing systems, therefore in some cases it is impossible to visit the entire neighborhood of a node. This kind of correlations cannot be easily included in the above mean-field analysis. On the other hand, Internet's local correlations may rapidly change in time, as they are affected by traffic congestions and routers' failures; therefore the average qualitative behavior of real processes should be close to the uncorrelated one.  \\
Writing $\tilde{R}_{2}(k,\ell) = \tilde{P}_{2}(k,\ell) / P(\ell)$, Eq.~\ref{P2} can be rewritten
\begin{equation}\label{P2bis}
\tilde{R}_{2}(k,\ell) = \sum_{m,n =1}^{\ell} \sum_{r =0}^{\ell}\mathcal{B}(\ell,m,n,r) \tilde{R}_{1}(n,\ell) \tilde{R}_{1}(m,\ell) \delta(k-n-m+r)~,
\end{equation}
and the merging process can be easily generalized to any number $\nu$ of sources exploiting the recursion relation, 
 \begin{equation}\label{Pnu}
\tilde{R}_{\nu}(k,\ell) =  \sum_{m,n =1}^{\ell} \sum_{r =0}^{\ell}\mathcal{B}(\ell,m,n,r)  \tilde{R}_{1}(n,\ell) \tilde{R}_{\nu -1}(m,\ell)  \delta(k-n-m+r)~.
\end{equation}
The degree distribution of the network obtained merging $\nu$ trees is then given by $\tilde{P}_{\nu}(k) = \sum_{\ell} P(\ell)\tilde{R}_{\nu}(k,\ell)$.
Note that Eq.~\ref{Pnu} is general and holds, within the validity of the approximation, for any type of process in the class under study, while the explicit expression of $\tilde{R}_{1}(k,\ell)$ as well as the correct form of $\mathcal{B}(\ell,m,n,r)$ depend on the details of the dynamics. \\
In the next section, we show how these methods can be applied to the traceroute model that describes the experiments used to determine the topology of the Internet. 

\begin{figure}[t]
\centerline{
\includegraphics*[width=0.5\textwidth]{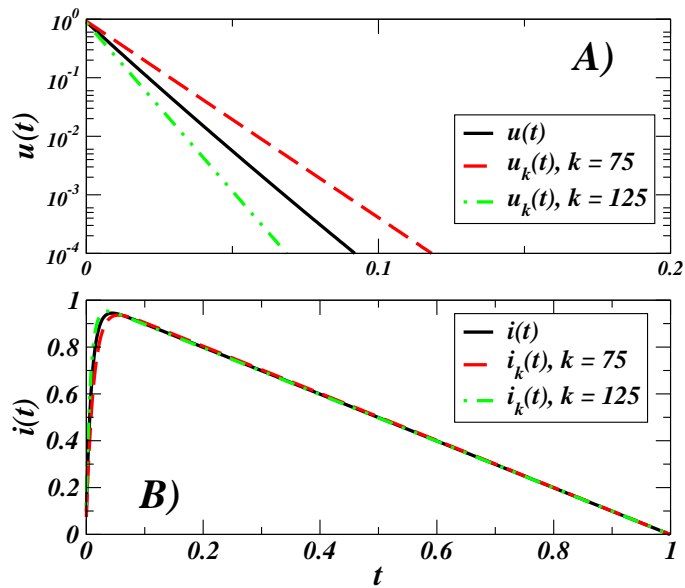}
}
\caption{ Temporal behavior of the global and partial densities of unreached (A) and interfacial (B) nodes in an homogeneous random graph with Poisson degree distribution of average degree $z=100$. The curves are obtained by numerical integration of the system in Eq.~\ref{eqnSTk}.  
  } 
\label{fig2}
\end{figure}

\section{Application to the Internet's mapping techniques}\label{sec3}

The Internet's topology can be studied at different levels. The most detailed Internet's descriptions are obtained at the level of single {\em routers}, but coarse grained representations are usually preferred for the possibility of obtaining a more reliable picture of the system (see Ref.~\cite{ves-book} for a simple introduction to the networked representations of the Internet). Routers sharing the same administrative policies are divided into {\em Autonomous Systems} (AS), that define the most important coarse-grained level of the Internet's topology. The first maps of the Internet were collected mainly at the AS level, using empirical data extracted from BGP tables together with those obtained by dynamical sampling methods based on traceroute measurements from single source \cite{pansiot,govindan,faloutsos}. 
According to these maps, that collect partial views of the net from some favored points, the Internet should be a very heterogeneous network with approximately power-law degree distribution $P(k) \propto k^{-\gamma}$, and $\gamma \simeq 2.1 \div 2.4$ \cite{faloutsos,govindan}. Because of the peculiar properties of scale-free networks, this discovery motivated a series of theoretical works in which toy-models of the Internet's mapping process were proposed and analysed in order to justify or question this empirical evidence \cite{lakhina,clauset1,latapy,dallasta}.\\
In a traceroute experiment, a given number of probes are sent from a source to a set of target nodes, tracing back the path followed during the exploration.
These probes are data packets that follow the same paths normally used by information to flow throughout the Internet. Although traffic congestions and local network's policies may cause unpredictable path's inflation, the traceroute paths are optimized in order to be the shortest ones between the source and the target nodes. Therefore, standard theoretical models of traceroute's explorations assume that the probes follow one of the possible shortest paths between the source and the destination. More precisely, we can include in the path only one shortest path among all equivalent ones (either a priori fixed or randomly chosen), or all of them.  
All strategies can occur in realistic processes, that is probably a mixture of them, but people usually give special attention to the one with a unique choice of the shortest path between nodes, that clearly brings to the worst overall sampling. We will also consider this case.\\ 
A one-to-all traceroute process is thus represented by an iterative algorithm running on a given network, that starting from a single source generates a spanning tree to all other nodes. Multi-source processes consist in merging different single-source spanning trees. In general, the reliability of traceroute-like sampling methods strongly depends on both the number of sources deployed on the network and the level of degree heterogeneity \cite{dallasta}. 
In fact, Lakhina et al.~\cite{lakhina} first showed numerically that sampling from single sources introduces uncontrolled biases and the observed statistical properties may sharply differ from the original ones.
More recently, Clauset and coworkers \cite{clauset1,clauset2} have pointed out that, because of the particular search procedure, a one-to-all traceroute tree has a power-law degree distribution $\tilde{P}_{1}(k) \propto k^{- \alpha}$ even if the underlying network is not scale-free. Actually, this was analytically proved only for homogeneous random graphs with fixed or Poisson degree distribution. In this case, the traceroute tree presents a power-law distribution with exponent $\alpha=1$ up to a cut-off equal to the average degree $z$. For networks with  power-law degree distributions $P(k) \propto k^{-\gamma}$, they suggested that the observed one should still be power-law but with a different exponent $\alpha < \gamma$. 
Within the same framework, but in partial contrast with this thesis, Cohen et al. \cite{cohen} have rigorously showed that in case of power-law networks, the bias on the exponent $\gamma$ is negligible. Other recent studies, based on mean-field approaches corroborated by numerical simulations, confirm the overall reliability of these mapping techniques on scale-free graphs \cite{dallasta,latapy}. Nevertheless, the debate on the traceroute process is still open, the main issues concerning the relevance of the biases in single-source processes and the improvements obtained using multiple sources. In the following, we try to address both these subjects using the theoretical approach developed in the previous section.\\
The analytical results mentioned above are actually based on approximated models that partially overlook the dynamical character of the process. The method used in Ref.~\cite{clauset1} to study one-to-all traceroutes is based on differential equations, but node sampling is essentially static. It assumes that a node of a given degree can appear with the same probability at any temporal step of the process. This is approximately true on homogeneous networks since the term $\frac{i_{k}(t)}{i(t)}$ in Eq.~\ref{P1} is $\simeq 1$, but it cannot be extended to the case of heterogeneous networks. 
The formal approach introduced in Ref.~\cite{clauset2} is more general, but it is still based on a uniformly random process, the so-called  ``exposure on the fly'' technique. It implicitly assumes a fitness-like variable homogeneously distributed on the ``stubs'' of a network, that plays the role of the time at which a node is explored. We improve this approximation using the dynamical method exposed in Section~\ref{sec2} for both single-source and multi-source processes. Obviously the framework becomes more complex and calculations can be performed analytically only in some special cases. 

\begin{figure}[t]
\centerline{
\includegraphics*[width=0.5\textwidth]{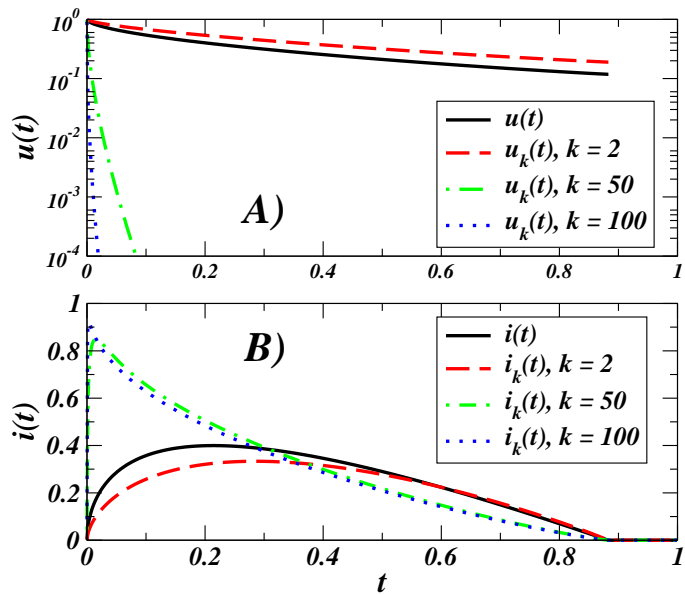}}
\caption{ Temporal behavior of the global and partial densities of unreached (A) and interfacial (B) nodes in a power-law random graph with exponent $\gamma = 2.5$ and average degree $z \simeq 4.5$. The curves are obtained by numerical integration of the system in Eq.~\ref{eqnSTk}. 
} 
\label{fig3}
\end{figure}

\subsection{Homogeneous Networks}
Let us consider the one-to-all traceroute exploration of an homogeneous random graph with Poisson degree distribution $P(k) = e^{-z} z^k / k!$. The original traceroute dynamics is discrete in time: at each temporal step, an interfacial node $\ell$ is randomly selected to spread out probes towards all its unknown neighbors; then all reached nodes are included in the interface, whereas node $\ell$ is moved to the bulk. A continuous-time approximation is defined grouping together $N$ discrete steps and passing to differential equations in the limit of large size $N$.\\
As the degree distribution is peaked around the average value $z$, one can safely do the further approximation that all nodes behave identically, that is  their temporal evolution is described by the mean-field densities $u(t)$, $i(t)$, $b(t)$ \cite{clauset1}. A selected interfacial node is connected to a unreached node with probability $p=z/N$, therefore in a temporal step, $p ~u(t)$ unreached nodes are visited and one node passes from the interface to the bulk. This process translates in the following system of equations \cite{clauset1} 
\begin{equation}\label{eqnST}
\left\{ \begin{array}{cl} 
\frac{d }{d t} u(t) & =  -z ~u(t) ~,\\
\quad & \quad\\
\frac{d}{d t}  i(t) & = + z ~u(t) - 1~, \\ 
\quad & \quad\\
\frac{d }{d t} b(t) & = + 1 ~.
\end{array} \right.
\end{equation}
The solution gives $u(t) = exp(- z t)$, $i(t) = 1- t -exp(- z t)$ and $b(t)=1-t$, with $t$ going from zero to a maximum value $T$, that is the first root of $i(t)=0$. Moreover, since in the traceroute sampling the probes emerging from the interfacial nodes visit all unreached neighbors, the mean-field probability to reach a node is just $\rho[u(t)] \simeq u(t)$.\\
The network's homogeneity implies that the internal degree profile of interfacial nodes is approximately the same of the underlying network at all times. According to this approximation,  for a traceroute spanning tree on an homogeneous Poisson graph Eq.~\ref{P1} reduces to
\begin{equation}\label{P1uncorr}
\tilde{P}_{1}(k+1) \simeq \sum_{\ell=k}^{\infty} \frac{1}{T} \int_{0}^{T} P(\ell+1) \left( \begin{array}{c} \ell \\ k  \end{array} \right) {\left[ e^{-z t} \right]}^{k} {\left[1 - e^{-z t} \right]}^{\ell -k} ~ dt~, 
\end{equation}
that is exactly the result obtained by Clauset and Moore \cite{clauset1}.\\
The integral in Eq.~\ref{P1uncorr} can be carried out noting that $T \simeq 1$ (for $z \gg 1$) and passing to the variable $u = u(t) = e^{-z t}$, with $d u = - z u dt$. Then, we can also easily perform the sum over $\ell$,  
\begin{equation}\label{P1uncorr-b}
\begin{split}
\tilde{P}_{1}(k+1) & \approx \sum_{\ell=k}^{\infty} P(\ell+1) \left( \begin{array}{c} \ell \\ k  \end{array} \right) {\left[ \frac{u^k}{z k} F_{2,1}(k,k-\ell,k+1,u) \right]}_{e^{-z}}^{1} \\
\quad &  \approx \frac{e^{-z}}{z k} \sum_{\ell=k}^{\infty} \frac{z^{l+1}}{l+1!} \left( \begin{array}{c} \ell \\ k  \end{array} \right) \left[ {\left( \begin{array}{c} \ell \\ k  \end{array} \right)}^{-1} - e^{-k z} F_{2,1}(k,k-\ell,k+1,e^{-z}) \right] \\
\quad & \approx \frac{1}{z k} \left[ 1 - \frac{z^k}{k!} e^{-z}\right] \simeq  \frac{1}{z k}~,
\end{split}
\end{equation}
where $F_{2,1}$ is the hypergeometric function, and the term proportional to $e^{-kz}$ is negligible for sufficiently large values of $k$. 
As in Ref.~\cite{clauset1}, we get an observed power-law degree distribution $\tilde{P}_{1}(k) \sim k^{-1}$ \cite{clauset1}, with a cut-off at $k \simeq z$. (It is worth to remark that Poisson random networks with average degree large enough to generate a power-law under sampling are extremely unlike in realistic systems, i.e. the Internet, community networks, etc. \cite{dallasta}.) \\
The results obtained simulating the traceroute model from a single source on a Poisson random graph (e.g. using the unique shortest path algorithm proposed in Ref.~\cite{dallasta}), confirm the $k^{-1}$ behavior of the observed degree distribution (see Fig.~\ref{fig1}-A).\\
In Fig~\ref{fig1}-A, we also report the observed degree distribution obtained sampling from two or more sources, that clearly depart from a purely power-law shape. 
The observed weird behavior, neither power-law nor poissonian, in which a peak appears at low degree values and moves forward for increasing number of sources, can be easily understood, at a qualitative level, using the theory presented in Section~\ref{sec2}. These peaks are due just to the superposition of power-law behaviors. In single-source experiments, most visited nodes have observed degree one, as they are discovered at the end of the process. The majority of them is rediscovered again in the same fashion during the second one-to-all process. Thus merging two spanning trees the overall distribution presents a peak at degree $2$ instead of $1$. The same happens for three sources, with a peak at degree $3$, and for increasing number of sources. In general, the position of the peak $k_{p}$ is not strictly equal to the number of sources $\nu$, but it usually hold $k_{p} \leq \nu$.
Solving numerically the recursive Eq.~\ref{Pnu}, we get the results reported in Fig.~\ref{fig1}-B. The curves have the same behavior as in the simulations, characterized by a peak at increasing degree values, then a decrease up to a cut-off about $k \approx z$.
Strikingly, the assumption of complete uncorrelation of successive spanning trees seems to be approximately correct for homogeneous random graphs. This is true up to a certain amount of sources (about $20$ in Fig.~\ref{fig1}-B), above which the ``metric'' correlations between shortest paths are not negligible. At this point, a peak at the original average degree $z$ is developed. \\
Our theoretical approach shows that the observed power-laws in one-to-all traceroutes on homogeneous random graphs are the result of a kind of convolution over a family of peaked symmetric distributions. This convolution process can be naturally ``inverted" by increasing the number of observation points. For a large number of sources the tree merging process corresponds to another kind of convolution on the power-laws emerging from single-source experiments and produces an unbiased sampling of the original degree distribution.
However, the minimal number $\nu^{*}$ of sources required to obtain an unbiased degree distribution is considerably large in homogeneous networks: from simple arguments and numerical evidences, $\nu^{*} \sim \mathcal{O}(z)$. \\   
In general the mean-field approximation on networks can be improved considering degree-dependent mean-field quantities, therefore we consider  the system of differential equations describing the dynamics for degree-dependent partial densities of bulk, interfacial and unreached nodes.
In uncorrelated networks, it reads 
\begin{equation}\label{eqnSTk}
\left\{ \begin{array}{cl} 
\frac{d }{d t} u_{k}(t) & =  - \sum_{h}\frac{(h-1)}{z} P(h) \frac{i_{h}(t)}{i(t)} k ~u_{k}(t) ~,\\
\quad & \quad\\
\frac{d}{d t}  i_{k}(t) & = + \sum_{h}\frac{(h-1)}{z} P(h) \frac{i_{h}(t)}{i(t)} k ~u_{k}(t)  - \frac{i_{k}(t)}{i(t)}~, \\ 
\quad & \quad\\
\frac{d }{d t} b_{k}(t) & = + \frac{i_{k}(t)}{i(t)} ~,
\end{array} \right.
\end{equation}
where $(h-1)\frac{i_{h}(t) P(h)}{i(t)} P(k|h)$ is the probability that emerging from an interfacial node of degree $h$ we reach a node of degree $k$ still unreached at time $t$.
Note that the above system reduces to Eqs.~\ref{eqnST} when the network is a regular random network of degree distribution $P(k) = \delta_{k,z}$. 
For poissonian networks, the exact time-depending behavior of the partial densities is quite complicated, as evidenced by the curves reported in Fig.~\ref{fig2} obtained solving numerically the system in Eq.~\ref{eqnSTk}. In the numerical solution we take initial conditions $i_{h}(0)=C_{0}\delta_{h,z}$ with $C_{0} \simeq \Delta t$, since both the initial condition and the temporal step $\Delta t$ should be of order $1/N$ in a system of size $N$. \\
The general behavior can be explained with simple arguments.
At the beginning of the process, the probability of having a node of degree $k$ at the interface is purely topological, i.e. $\frac{i_{k}(0^+)}{i(0^+)} \simeq \frac{k}{z}$. The late times behavior for $t \gg 0$, can instead be computed knowing that the corresponding  behavior of $i(t)$ is approximately linear, i.e. $i(t) \simeq 1-t$, and that $u_{k}(t)$ decreases exponentially fast in time. These results, obtained plugging the short times approximation $\frac{i_{k}(t)}{i(t)} \simeq \frac{k}{z}$ into the equation for $\frac{d u_{k}(t)}{d t}$, are verified in the numerics.  Hence, from Eq.~\ref{eqnSTk}, $\frac{d i_{k}(t)}{d t} \approx k e^{-k t} - \frac{i_{k}(t)}{1-t}$.
For sufficiently large $k$, the first term at the r.h.s. can be neglected, thus after integration we get $i_{k}(t) \simeq 1-t$.
Therefore, the quantity $\frac{i_{k}(t)}{i(t)}$ is expected to approach the unity for sufficiently large $t \gg t_{k}^{*}$, where $t^{*}_{k}$ is the time at which the maximum value of $i_{k}(t)$ is reached. From the previous arguments one expects $t^{*}_{k} \propto 1/k$.\\
These simple calculations, and the numerical results reported in Fig.~\ref{fig2}, show that the short time behavior of the sampling process is not trivial at all.  Nonetheless, the degree distributions obtained solving numerically the system in Eq.~\ref{eqnSTk} and plugging the corresponding partial densities in Eqs.~\ref{P1} and \ref{Pnu} are in perfect agreement with the analytical results based on the approximation of complete homogeneity (not shown).

\begin{figure}[t]
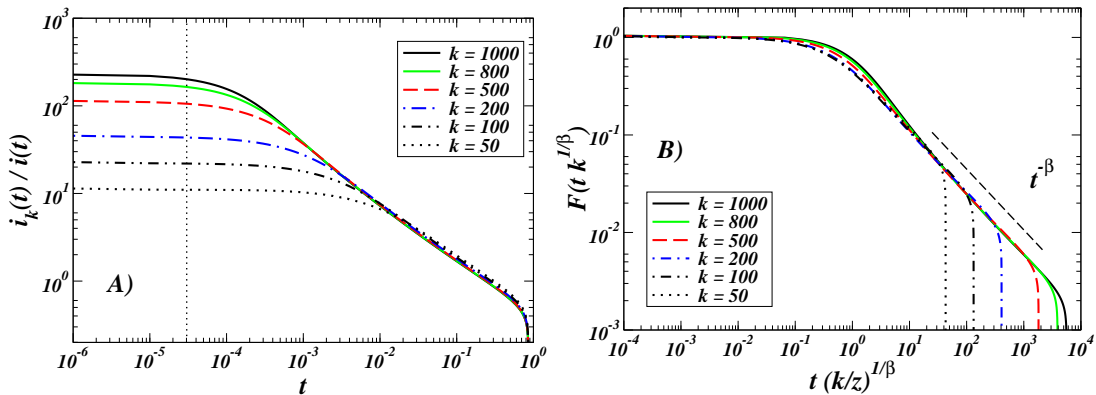

\centerline{
\includegraphics*[width=0.4\textwidth]{fig3B}
\includegraphics*[width=0.4\textwidth]{fig3C}
}
\caption{ A) Temporal behavior of the function $i_{k}(t)/i(t)$ for high-degree nodes in a power-law random graph. B) The scaling function $\mathcal{F}$ (see Eq.~\ref{scaling}) associated with some universal behavior of $i_{k}(t)/i(t)$ for high-degree nodes. 
} 
\label{fig3b}
\end{figure}

\subsection{Heterogeneous Networks}
\begin{figure}[t]
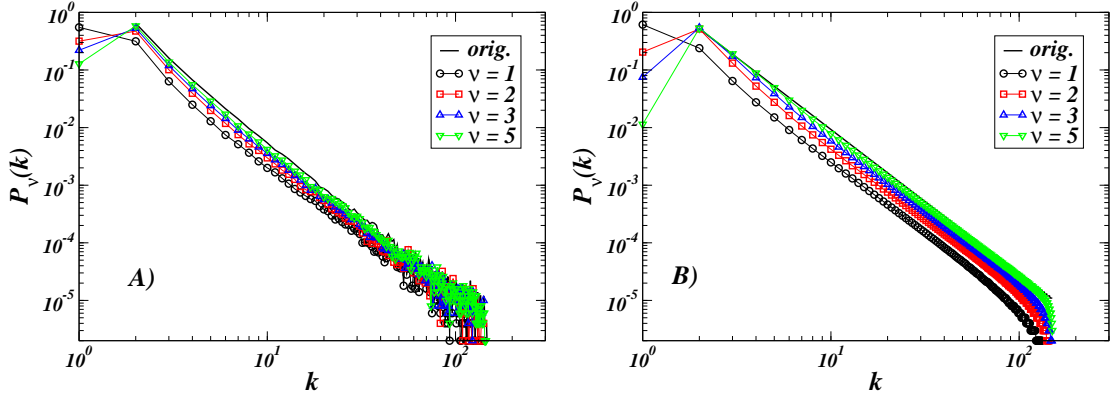

\centerline{
\includegraphics*[width=0.4\textwidth]{fig4A}~~
\includegraphics*[width=0.4\textwidth]{fig4B}
}
\caption{ A) Degree distribution $\tilde{P}_{\nu}(k)$ of the network produced by merging together $\nu$ spanning trees generated by one-to-all traceroutes algorithms on a power-law random graph of size $N=10^4$, average degree $z\simeq 4.5$ and exponent $\gamma \simeq 2.5$. The level of sampling, almost unbiased with just a single source, improves increasing the number of sources. B) Degree distribution $\tilde{P}_{\nu}(k)$ obtained numerically from the recursion relation in Eq.~\ref{Pnu} for a power-law random graph with average degree $z\simeq 4.5$ and exponent $\gamma \simeq 2.5$.
} 
\label{fig4}
\end{figure}

The degree inhomogeneity is instead expected to play an important role in the exploration of networks with skew and fat-tailed degree distributions, in which the dynamical sampling of nodes is far from being a uniform process. Though to obtain the exact form of partial and global densities it is necessary to solve the evolution equations explicitly, that is in general very difficult, the qualitative behavior of these quantities for large degree values ($\ell \gg z$) can be deduced with some approximate argument.
We will show that the tail of the degree distribution of a power-law random graph is sampled with negligible bias, even in single-source experiments.
This is mainly due to the fact that high degree nodes arrive at the interface of the process almost immediately, and their neighbors are fairly sampled with a probability that depends only weakly on the dynamics. \\ 
At the beginning of the process, nodes with large degree are preferentially sampled, i.e. $i_{\ell}(t) /i(t) \simeq \ell/z$, implying that the number of unreached hubs rapidly decays to zero. At this point, almost all high degree nodes are at the interface of the process, $i_{\ell}(t)$ varies slowly, and the sampling among high degree nodes inside the interface becomes almost uniformly random (i.e. hubs of degree $\ell$ are picked with probability $P(\ell)$). Hence, for sufficiently large times and degrees $\ell \gg z$, the quantity $\frac{i_{\ell}(t)}{i(t)}$ is expected to become independent of the degree $\ell$. Solving numerically Eqs.~\ref{eqnSTk}, we find that, for $t \gg 0$, $ \frac{i_{\ell}(t)}{i(t)} \propto {t}^{-\beta}$, with a cut-off close to the final sampling time $T$ and the exponent $\beta$ depending on the details of the degree distribution (not only on $\gamma$). 
The numerical results presented in Fig.~\ref{fig3b} suggest a scaling function for $\frac{i_{\ell}(t)}{i(t)}$ of the type 
\begin{equation}\label{scaling}
 \frac{i_{\ell}(t)}{i(t)} \approx \frac{\ell}{z} ~ \mathcal{F} \left[ t {\left( \frac{\ell}{z}\right)}^{1/\beta}\right]~,
\end{equation} 
with $\mathcal{F}(x) \propto x^{-\beta}$ when $x \gg 1$ and $\mathcal{F}(x) \approx 1$ when $x \ll 1$. The scaling form is correct up to a time $T_{\ell} \sim T {(\ell/z)}^{-1/\beta}$, at which the quantity vanishes. Note that for a system of size $N$, and power-law degree distribution of exponent $\gamma$, the maximum degree scales as $\ell_{max} \sim N^{1/(\gamma-1)}$. Since the temporal step of the dynamics (integration step) is $\Delta t \sim 1/N$, a realistic minimum observation time for the above curves is ${\Delta t}^{1/(1-\gamma)}$, that is traced in Fig.~\ref{fig3b}-A with a dotted vertical line. It is clear that for high-degree nodes the power-law scaling behavior dominates the most of the dynamics. The result is surprising since one would expect that, in a power-law network, $i_{\ell}(t) /i(t) \simeq \ell/z$ during the whole dynamics. We will see that this behavior is crucial to get an unbiased estimate of the exponent. \\
The other interesting time-dependent quantity is the probability to visit an unreached node, $\bar{u}(t) = \sum_{\ell} \frac{\ell}{z} P(\ell) u_{\ell}(t)$. At long times and high degrees $\ell$, the quantity $u_{\ell}(t)$ is non zero only for low degree nodes, so the temporal behavior of $\bar{u}(t)$ is similar to that of $u(t)$. Actually, it seems to decay from $1$ with a law that is clearly slower than an exponential one, but  faster than a power-law. Indeed, if $i_{\ell}(t)/i(t)$ follows a power-law behavior, formally integrating Eqs.~\ref{eqnSTk} one gets $\bar{u}(t) \sim e^{-a t^{\alpha}}$, with $\alpha <1$. However, numerical integration does not clarify the possible relation existing between $\alpha$ and $\beta$. \\
Plugging the above results into the sampling integral (Eq.~\ref{P1}), it is actually possible to show numerically that the observed degree distribution maintains the same functional form of the original one, at least for high degree nodes. The comparison between observed degree distribution obtained from simulations and by solving numerically the sampling equations are reported in Fig.~\ref{fig4}. The qualitative behavior is the same: the traceroute sampling on power-law random graphs reproduces the original degree distribution without any significative bias.\\
Some further insights on the reason of this result can be obtained with the following rough argument.
Let us consider the sampling formula 
\begin{equation}\label{P1STk}
\tilde{P}_{1}(k+1)  \simeq  \sum_{\ell=k}^{\infty} \frac{1}{T} \int_{0}^{T} \frac{i_{\ell+1}(t) P(\ell+1)}{i(t)} \left( \begin{array}{c} \ell \\ k  \end{array} \right) {\left[ \bar{u}(t) \right]}^{k} {\left[1 - \bar{u}(t) \right]}^{\ell -k} ~ dt 
\end{equation}
and approximate it for $\ell \gg z$ considering that 1) during most of the dynamics the quantity $i_{k}(t)/i(t)$ assumes the scaling form $t^{-\beta}$, and 2) the binomial probability can be approximated by a gaussian peaked around its maximum, i.e. $\bar{u}^{*} \sim k/\ell$. 
A rough estimate can be done performing the gaussian integral at the saddle-point or, in an equivalent way, recalling a property of Dirac delta functions, i.e.  $\int f(t) \delta(g(t)) dt \approx \sum_{i} f(t_{i})/|g'(t_{i})|$, where $t_{i}$ are the zeros of $g(t)$ and $g'(t_{i})$ is the derivative of the function $g$ in these points. For large $k$ and $\ell$, $g(t) \approx \ell \bar{u}(t) - k$, thus $|g'(t)| \propto \ell {|\frac{d \bar{u}}{dt}|}_{t_{\ell}}$ (in which $t_{\ell}$ is the time at which $\bar{u}(t_{\ell}) \simeq u^{*} \simeq k/\ell$ for given values of $k$ and $\ell$). 
Putting all ingredients together, the sampling formula becomes
\begin{equation}\label{P1new}
\tilde{P}_{1}(k+1)  \simeq  \sum_{\ell=k}^{\infty} P(\ell+1) \frac{1}{\ell+1}\frac{t_{\ell}^{-\beta}}{{|\frac{d \bar{u}}{dt}|}_{t_{\ell}}}~, 
\end{equation} 
Since $\bar{u}(t) \sim e^{-a t^{\alpha}}$, the derivative is proportional to $k/\ell$ times some power-law in time, we get 
\begin{equation}\label{P1new2}
\tilde{P}_{1}(k+1)  \propto  \sum_{\ell=k}^{\infty} P(\ell+1) \frac{1}{k} {\left[\log(\frac{\ell}{k})\right]}^{\frac{1-\beta-\alpha}{\alpha}}~.
\end{equation}
For large degree we can neglect the logarithmic contributions in the sum over $\ell$, finding $\tilde{P}_{1}(k)  \propto  k^{-\gamma}$. \\
According to this result, the degree distribution of the spanning tree emerging from one-to-all traceroute samplings of a scale-free graph (with exponent between $2$ and $3$) is qualitatively the same of that of the underlying network, at least for high degree nodes. 
The result is in agreement with a recent analysis by Cohen and coworkers \cite{cohen}, in which some rigorous bounds for the traceroute biases in power-law networks are obtained using the exposure technique. We believe that the general picture describing the behavior of traceroute-like processes could be extracted from the present approach in a much easier way than from the ``exposure on the fly'' method. Moreover, the framework can be straightforwardly extended to study, at least numerically, the effects of other relevant parameters, and correlations.  In these perspective our approach can be considered as a benchmark that could be useful in the problem of network inference \cite{viger} and bias reduction \cite{flaxman}. 

\section{Conclusions}\label{sec4}

The study of the interplay between topological and dynamical properties of networks is of primary interest in the current research on complex systems.
With the present work, we introduce in this framework a general method to investigate the topological properties of growing clusters that are dynamically defined by a given class of spreading processes and algorithms. These are processes that starting from a single source span the whole network, traversing all nodes only once. The temporal evolution is defined by means of a system of differential equations for the (partial) densities of bulk, interfacial, and unreached nodes. \\ 
Our approach, that allows to compute analytically or numerically the degree distribution of the emerging tree-like structure, is based on the idea that during its evolution the dynamics performs a sampling of the local structure of the underlying network. As the sampling rate depends on the dynamical properties themselves, the degree distribution of the emerging subnetwork may differ considerably from that of the original network. 
The generalization to study multiple-source processes is also discussed, at least in the approximation that the overall structure is obtained merging together collections of independent single-source processes. \\
A natural application of the method presented here consists in the analysis of the dynamical sampling of networks. We have provided 
a deeper insight in the qualitative behavior of traceroute-like processes, shedding light on the dynamical mechanism at the origin of the
observed topology. The reasons for the different sampling accuracy observed in homogeneous and heterogeneous networks should be now clear. In particular, we have shown that on homogeneous networks,
\begin{itemize}
\item in a single source sampling, the network is observed to have a power-law degree distribution with exponent $-1$ up to a cut-off about the original average degree $z$; 
\item increasing the number of sources destroys the power-law behavior, but a fair sampling requires about $\mathcal{O}(z)$ sources;
\item ``metric'' correlations (that can be associated to the betweenness centrality of nodes) favor a better sampling of the local topology for increasing number of sources.
\end{itemize}
In contrast, on heterogeneous networks, 
\begin{itemize}
\item high-degree nodes are preferentially sampled at the beginning of the process, and with higher accuracy (in this case high-degree nodes are essentially those with highest betweenness);
\item in single-source experiments, power-law degree distributions are sampled with negligible biases;
\item the overall sampling of the degree distribution becomes very accurate with just few sources.  
\end{itemize}
We believe that a good understanding of these kind of processes is fundamental in order to improve the performances of current dynamical sampling techniques applied to technological networks such as the Internet and the Web.\\
As mentioned in the Introduction, there are other dynamical processes that can be described using the present approach, from the epidemic spreading, to broadcast trees and search techniques. We just mention a couple of cases that may find further developments. One consists in a recently proposed model of search in social networks, in which the search efficiency decays with the distance \cite{adamic}. Let us consider an uncorrelated homogeneous random network, on which we perform a distance dependent snowball search such that the probability to visit a still unreached neighbor of an interfacial node is $\theta(t) \propto  {(t+A)}^{-\beta}$, with $\beta \in [0,1)$. The corresponding global density is obtained solving the equation $\frac{d u(t)}{dt} = - z \theta(t) u(t)$. 
The limit in which $\beta \to 1$ is particularly instructive, since the global density decreases much slower than an exponential. Performing the calculations, $u(t) \sim A/(A+t)$ and the emerging tree presents a degree distribution $\tilde{P}_{1}(k) \approx k^{-2}$ up to a cut-off around $z$. The example shows how easy can be to find processes that generate power-law degree distributions out of exponential networks.\\
The second relevant example concerns epidemic-like spreading phenomena. Here the system of differential equations governing the dynamics is the celebrated susceptible-infected-removed (SIR) model \cite{sir}, in which unreached nodes are identified with susceptible nodes, interfacial nodes with infected ones and the bulk nodes correspond to removed individuals. In analogy with the traceroute dynamics, we can write a system of first-order differential equations governing the temporal evolution of partial densities \cite{romu}. The fundamental difference between the two dynamics is that in the traceroute model at each time step the algorithm chooses an interfacial node to sample its neighbors, while in the SIR model all infected nodes have a fixed probability to spread the virus to their neighbors (with spreading rate $\lambda$). However, for small values of $\lambda$ (above the percolation threshold) the topological structure is still tree-like and can be analyzed as in Section~\ref{sec2}. 
When different transmission properties (e.g. degree-dependent and distance-dependent rates $\lambda$) are taken into account, the growing infection region might display very non-trivial topologies. At the same time, the knowledge of the topological structure of an infection's outbreak may be relevant for the design of more appropriate immunization strategies.\\
From a general point of view, the present formalism can be extended to study (at least numerically) the effect of degree-degree correlations or quenched disorder, that have not been considered here but play an important role in all real experiments. 

In conclusion, we expect that analyses like the one performed in this paper will allow to get a better understanding of the functional interplay between a network and the dynamical processes evolving on it.

\begin{acknowledgements}
The author is grateful to A. Barrat, G. Bianconi and M. Marsili for fruitful and stimulating discussions, and to D. Begh\'e for the constant encouragement during this work. 
\end{acknowledgements}

\end{document}